\documentclass[aps,pra,twocolumn,superscriptaddress]{revtex4-1}

\usepackage{amsmath}
\usepackage{amssymb}
\usepackage{graphicx}
\usepackage{dcolumn}
\usepackage[colorlinks=true, allcolors=blue]{hyperref}
\usepackage[super]{nth}
\usepackage{blindtext}
\usepackage{siunitx}

\usepackage{textgreek}
\usepackage{float}

\usepackage{comment}
\begin{document}

\title{Measurement of the $2S_{1/2},F$=$0 \rightarrow 2P_{1/2},F$=$1$ transition in Muonium}

\author{G.~Janka}
\author{B.~Ohayon}
\author{I.~Cortinovis}
\author{Z.~Burkley}
\author{L.~de Sousa Borges}
\author{E.~Depero}
\affiliation{
Institute for Particle Physics and Astrophysics, ETH Z\"urich, CH-8093 Z\"urich, Switzerland
}
\author{A.~Golovizin}
\affiliation{P.N. Lebedev Physical Institute, 53 Leninsky prospekt., Moscow 119991, Russia
}
\author{X.~Ni}
\author{Z.~Salman}
\author{A.~Suter}
\affiliation{Laboratory for Muon Spin Spectroscopy, Paul Scherrer Institute, CH-5232 Villigen PSI, Switzerland}
\author{T.~Prokscha}
\affiliation{Laboratory for Muon Spin Spectroscopy, Paul Scherrer Institute, CH-5232 Villigen PSI, Switzerland}
\author{P.~Crivelli}\email{Corresponding author: crivelli@phys.ethz.ch}
\affiliation{
Institute for Particle Physics and Astrophysics, ETH Z\"urich, CH-8093 Z\"urich, Switzerland
} 
\collaboration{Mu-MASS collaboration}

\date{\today}

\begin{abstract}
Muons are puzzling physicists since their discovery when they were first thought to be the meson predicted by Yukawa to mediate the strong force. The recent results at Fermilab on the muon g-2 anomaly puts the muonic sector once more under the spotlight and calls for new measurements with this fascinating particle.
Here we present the results of the first measurement of the $2S_{1/2},F$=$0 \rightarrow 2P_{1/2},F$=$1$ transition in Muonium, the hydrogen-like bound state of a positive muon and an electron. The measured value of \SI{580.6+-6.8}{\mega\hertz} is in agreement with the theoretical calculations. From this measurement a value of the Lamb shift of \SI{1045.5+-6.8}{\mega\hertz} is extracted, compatible with previous experiments. We also determine for the first time the $2S$ hyperfine splitting in Muonium to be \SI{559.6+-7.2}{\mega\hertz}. 
The measured transition being isolated from the other hyperfine levels holds the promise to provide an improved determination of the Muonium Lamb shift at a level where bound state QED recoil corrections not accessible in hydrogen could be tested. Such a measurement will also be sensitive to new physics in the muonic sector, e.g. to new bosons which might provide an explanation of the g-2 muon anomaly or Lorentz and CPT violation. We also present the first observation of Muonium in the $n = 3$ excited state opening up the possibility of new precise microwave measurements as realized in hydrogen.  

\end{abstract}

 
\maketitle

\section{Introduction}
The discovery of muons has an intriguing history. As in the case of the positron, they were first detected in cosmic radiation by Anderson and Neddermayer in \num{1936} \cite{Anderson1937}. Interestingly, the first hints of muons were actually already seen in \num{1933} by Kunze \cite{Kunze} in his Wilson chamber, but he was not confident enough about his results to claim the discovery of a new particle. 

The muon was first thought to be the pion predicted by Yukawa (1935) \cite{Yukawa}, the heavy quantum (meson) responsible of mediating the nuclear (strong) force in analogy to the light quantum (photon) for the electromagnetic interaction. The pion based on the range of the nuclear force should have had a mass of \numrange{100
}{200} times the mass of the electron and should be both positively and negatively charged. The muon (\num{207} times the electron mass) seemed to have just the expected value. However, in \num{1946} an experiment of Conversi et al.~\cite{Conversi1946} showed that their interaction with nuclei was too weak to be attributed to pions. They observed that the negative \textit{mesotron} would decay instead of being absorbed by carbon after having formed a pionic atom as predicted by Tomonaga-Araki \cite{Tomonaga}. This was the first indication of the formation of muonic atoms as it was realized a few years later. Moreover, the lifetime of the muon was of the order of $10^{-6}$ \si{\second} which is $10^{12}$ times longer than expected by strong interaction processes. Finally in \num{1947}, Powell et al.~\cite{Powell} detected the pion using photographic emulsions at high altitudes and verified that it decays into a muon and a neutral particle that was identified to be a neutrino. The possibility that this could be a photon was eliminated by the non-observation of pair production as expected in this case. This led Isidor Rabi to come up with his famous quote: “Who ordered that?”. 

The fascinating history of the muon continues to this day. The recent results at Fermilab \cite{2021_Abi} confirming that the measured muon anomalous magnetic moment (g-2 muon) deviates from the standard model prediction by \num{4.2} standard deviations calls for further scrutiny. Muonium (M), the bound state of the positive muon ($\mu^+$) and an electron is an ideal system to study the muon properties and hunt for possible new effects. Due to its lack of sub-structure it is free from finite size effects and is therefore an excellent candidate to probe bound-state QED \cite{2005_Karshenboim} and search for new physics beyond the Standard Model \cite{2014_Karshenboim, 2014_Vargas, 2019_Dark}. 

Precise measurements of the ground-state hyperfine structure (HFS) \cite{1999_Liu} and the $1S-2S$ transition \cite{2000_Meyer} in Muonium were performed, with improvements proposed by MuSEUM (HFS, \cite{2020_MUSEUM}) and Mu-MASS ($1S-2S$, \cite{2018_Crivelli}) ongoing. All the measurements so far of the M Lamb shift (LS) are limited by statistics \cite{1984_Oram,1990_Woodle, 1990_Kettell}. Only recently, the formation of an intense metastable M(2S) beam was demonstrated by the Mu-MASS collaboration at the low-energy muon beamline (LEM) at PSI \cite{2020_Janka}, opening up the possibility for a new generation of precision measurements of the M LS. 

\begin{figure}[!b]
\includegraphics[width=1\columnwidth,trim={0 0 0 60},clip]{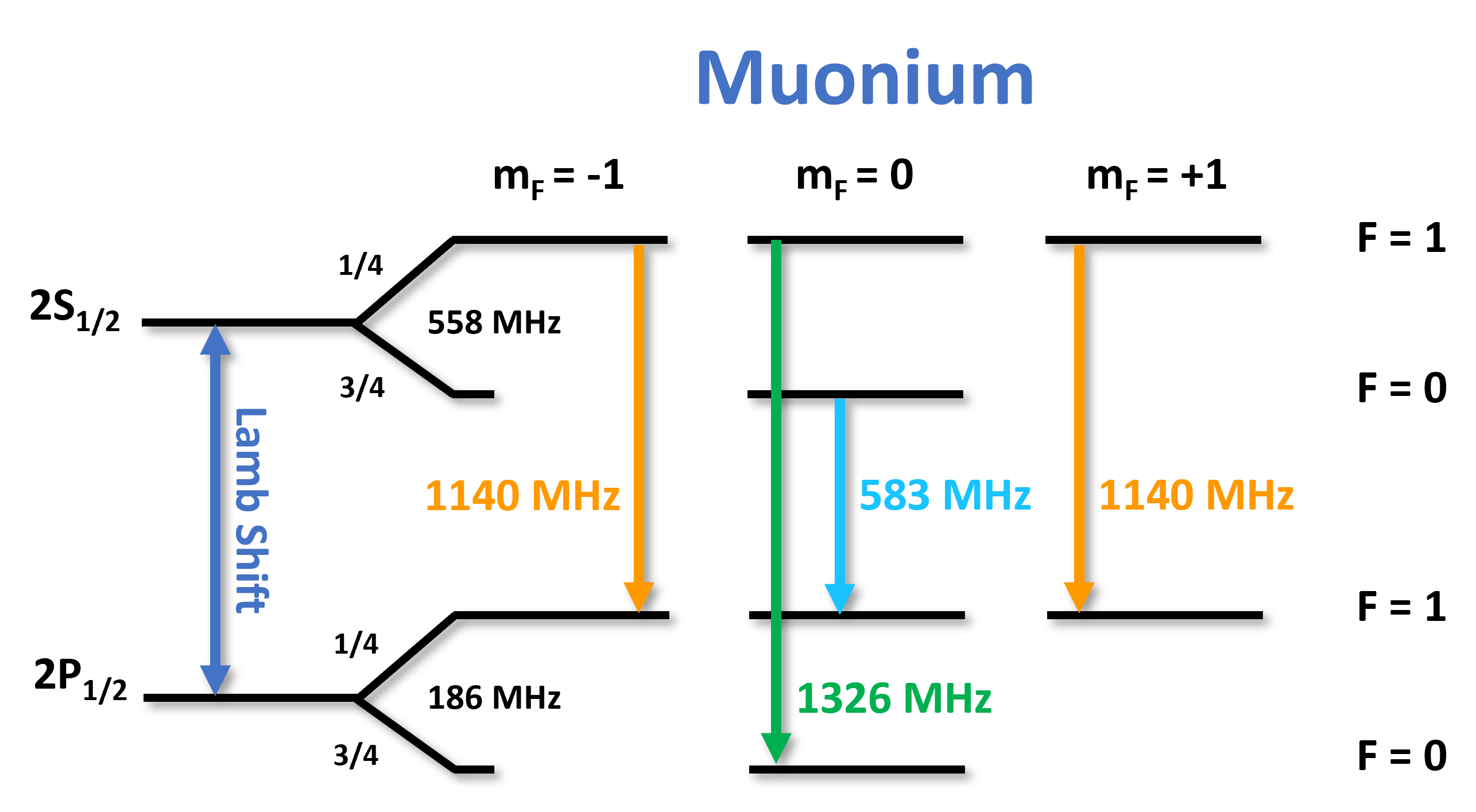}
\caption{\label{fig:mu_transitions} The hyperfine transitions of the M $2S_{1/2}-2P_{1/2}$ levels, including the $2S$ and $2P$ hyperfine splittings.}
\end{figure}

\begin{figure*}[!t]
\centering
\includegraphics[width=2\columnwidth,trim={0 0 0 0},clip]{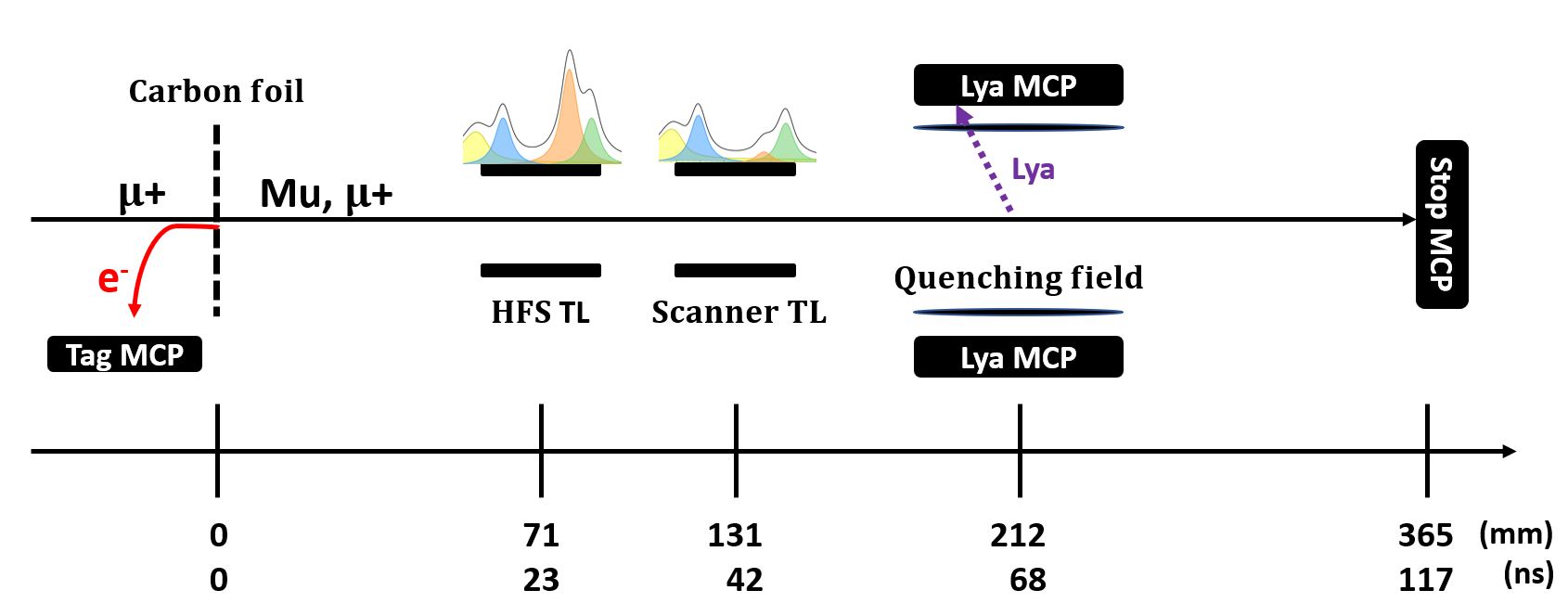}
\caption{\label{fig:beamline} 
Schematic view of the experimental setup. The signal signature consists of a coincidence signal between Tag,  Ly$\alpha$ and Stop-MCP. The coloured line-shapes above the TLs correspond to the observable transitions (\SI{1326}{\mega\hertz} (green), \SI{1140}{\mega\hertz} (orange), \SI{583}{\mega\hertz} (blue) and the combined $3S_{1/2}-3P_{1/2}$ (yellow) contribution). The time scale is given for the most-probable Muonium atom with an energy of \SI{5.7}{\kilo\electronvolt}.}
\end{figure*}

A first measurement at the LEM beamline resulted in a M LS value of \SI{1047.2+-2.5}{\mega\hertz} \cite{2021_Ohayon}, representing an order of magnitude improvement upon the previous measurements. With the muCool beamline \cite{2021_antognini} and (if approved) the high intensity muon beam (HiMB) upgrade at PSI \cite{2021_HiMB2}, the $\mu^+$ beam quality and flux will further improve, making it feasible to reach statistical uncertainties of the order of a few tens of \si{\kilo\hertz} within a few days of beamtime. This would allow to probe QED corrections enhanced in the Muonium system such as the Barker-Glover effect (\SI{160}{\kilo\hertz}) or the nucleus self-energy (\SI{40}{\kilo\hertz}), which are still not in reach for hydrogen \cite{2021_Janka}.

The prospect of not being limited by statistics calls for a systematically more robust method to extract the LS. Following the example of the most precise measurements of the hydrogen Lamb shift \cite{1979_Newton, 1986_Lundeen, 2019_Hessels}, extracting the LS from the isolated hyperfine transition $2S_{1/2},F$=$0\rightarrow 2P_{1/2},F$=$1$ is preferred over the other two allowed transitions (see Fig.~\ref{fig:mu_transitions}). Due to the reduced influence of other nearby transitions, the systematic uncertainties related to line-pulling would become negligible.

We report here the first measurement of the isolated hyperfine transition, from which we extract the M LS. Combining this result with our previous measurement \cite{2021_Ohayon}, we determine the $2S$ hyperfine splitting in Muonium for the first time. Additionally, we present the first observation of Muonium in the $n=3$ excited state.

\section{Experimental Method}
The LEM beamline at PSI provides low (1-30 keV) energy muons with a rate of up to \SI{20}{\kilo\hertz} using a solid neon moderator \cite{2008_uE4}. In our experiment, we set the energy of the $\mu^+$ to \SI{7.5}{\kilo\electronvolt} in order to maximize the number of M(2S) available for the measurement \cite{2020_Janka}. The beam transport is optimized with several lenses along the beamline, eventually focusing the muons onto a carbon foil at the entrance of the setup shown in Fig.~\ref{fig:beamline}. 

In contrast to the LS measurements with protons, the flux of muons is around ten orders of magnitude smaller, which makes it essential to have an efficient background suppression without sacrificing efficiency. Therefore, the incoming muons are tagged on an event by event basis to be able to discriminate their times-of-flight (TOFs).
When impinging onto the \SI{10}{\nano\meter} thin carbon foil, a $\mu^+$ releases secondary electrons. These electrons are detected by a microchannel plate (Tag MCP) to give the start signal of the TOF measurement. Upon leaving the foil, the tagged $\mu^+$ has a probability of picking up an electron and form M, primarily in the ground state. From measurements with hydrogen \cite{1981_Gabrielse, 1973_Tielert}, \SIrange{5}{10}{\percent} is expected to be formed in $n=2$ and \SI{2}{\percent} in $n=3$. This expectation was confirmed with a measurement of \SI{11+-4}{\percent} \cite{2020_Janka} for M in the metastable state. The M(2S) lifetime is limited by the decay time of the muon itself ($\tau_\text{M(2S)}$=\SI{2.2}{\micro\second}). M(3S) is unstable ($\tau_\text{M(3S)}$=\SI{158}{\nano\second}) and will relax back to the ground state via the intermediate $2P$ state ($\tau_\text{M(2P)}$=\SI{1.6}{\nano\second}).

The formed beam passes first through a transmission line (TL) which depopulates unwanted hyperfine states (HFS TL). As established by the most precise H LS measurements, this reduces the background and at the same time narrows and simplifies the overall line-shape to be fitted. By setting the HFS TL at a frequency of \SI{1140}{\mega\hertz} and with an average power of roughly \SI{29}{\watt}, according to our simulation, we depopulate the $F\mathrm{=}1, m_F\mathrm{=}\pm 1$ states by \SI{88}{\percent} and the $F\mathrm{=}1, m_F\mathrm{=}0$ state by \SI{16}{\percent}. The transition of interest $2S_{1/2},F$=$0 \rightarrow 2P_{1/2},F$=$1$ at \SI{583}{\mega\hertz} remains almost unaffected, where its initial state $F\mathrm{=}0, m_F\mathrm{=}0$ is only depopulated by \SI{4}{\percent}. With a second TL (Scanner TL) we scan the line-shape of this transition. Seven frequency points are measured in the range of \SIrange{200}{800}{\mega\hertz} with an average power of \SI{22.3}{\watt}. An additional data point is taken with the Scanner TL turned off. The output powers of both TLs are continuously monitored with power meters to ensure the stability of the measurement.

Due to the short lifetime of the $2P$ states, the atoms excited by the microwave relax to the ground state before reaching the detection setup. The remaining excited states are quenched by applying an electrical field of the order of  $\SI{250}{\volt\per\centi\meter}$ between the two grids mounted in front of the coated Ly$\alpha$-MCPs (see Fig.\ref{fig:beamline}) and relax back to the ground state with the emission of UV photons (2P Ly$\alpha$: \SI{121.5}{\nano\meter}, 3P Ly$\beta$: \SI{102.5}{\nano\meter}). These photons interact with the coating and release single electrons, which are in turn detected by the Ly$\alpha$-MCPs. The beam then continues onto the Stop-MCP, giving the stop time for the TOF measurement.
An event is recorded only if a coincidence signal between the Tag- and Stop-MCP is seen above threshold (double coincidence). The time window for a valid event is set to \SI{10}{\micro\second} and multiple hits in the detectors are recorded. 
The signature for the detection of M 2S atoms is defined as the double coincidence and additionally the Ly$\alpha$-MCP signal in the time region of interest (triple coincidence, see also Fig. \ref{fig:beamline}). 

\begin{figure}[h!]
    \includegraphics[width=1\columnwidth,trim={14 12 20 22},clip]{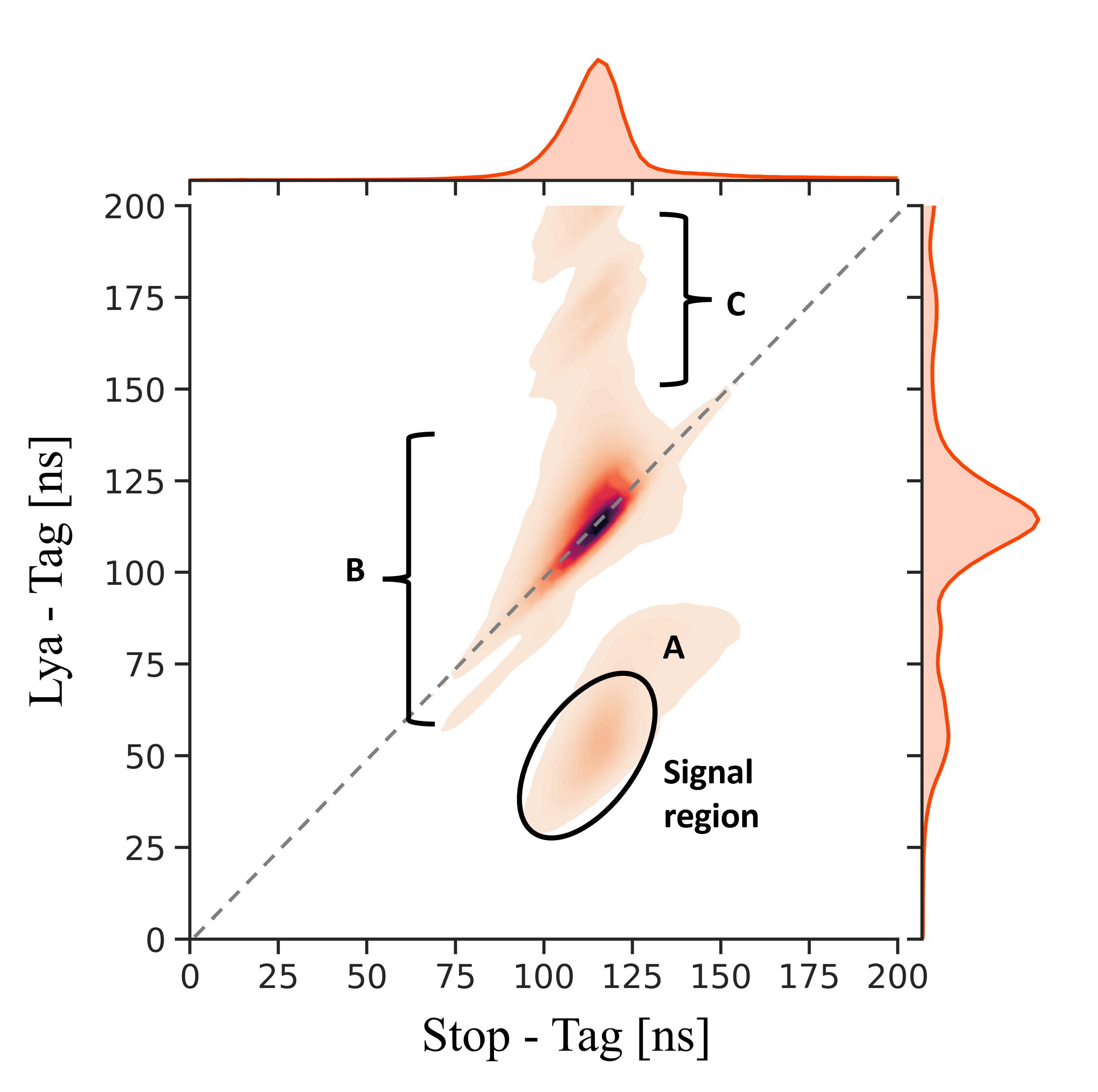}
    \caption{ \label{fig:LS_cuts} Correlation plot of the TOF between the Tag- and Stop-MCP and the TOF between the Tag- and Ly$\alpha$-MCP. The signal region of Ly$\alpha$ events is circled. The regions labelled A,B,C are explained in the main text.}
\end{figure}

In Fig.~\ref{fig:LS_cuts}, the measured TOFs between Tag- and Stop-MCP (x-axis) are correlated with the TOFs between Tag- and Ly$\alpha$-MCP (y-axis). The energy distribution of the M atoms after the carbon foil is well-known from previous measurements at the LEM beamline and reproduced by simulation \cite{2015_Khaw}. The most-probable energy of the $\mu^+$ after the foil is \SI{5.7+-0.2}{\kilo\electronvolt}. The expected TOF of an M atom from the Tag- to the Stop-MCP ranges from \SIrange{90}{135}{\nano\second}, with a most-probable TOF of \SI{117}{\nano\second}. Applying this time cut, we extract the detected amount of M and $\mu^+$, which serves as normalization factor $S_\text{Norm}=N_M + N_{\mu^+}$, accounting for variations in the beam intensity. 
The TOF between the tagging and the emission of a Ly$\alpha$ photon is calculated to be in the range of \SIrange{30}{75}{\nano\second}. Applying both of these cuts, we identify our signal region, portrayed in Fig.~\ref{fig:LS_cuts}. The amount of signal events are extracted for each frequency point ($S_\text{Lya}(f)$). The normalized signals per frequency point are then calculated as $S(f)=S_\text{Lya}(f)/S_\text{Norm}$.

In region A of Fig.~\ref{fig:LS_cuts}, events with longer TOFs than expected are summarized. Those events might be caused by either convoy electrons created at the foil \cite{1986_Focke} or by ionization products of a $\mu^+$ interacting with the residual gas. The region A is leaking into the signal region, contributing to the measured background. 
The region B contains the events where the TOFs between the Ly$\alpha$- and Stop-MCP are almost zero. These MCPs are close to each other, which leads to the risk of one detector picking up a noise signal when the other one is firing or vice versa. This noise signal can have a large amplitude with an intense after-ringing, which is mistaken as an additional signal that contributes to the events in region C. 

The average power inside the TLs is frequency-dependent. To be able to fit a line-shape, all frequency points are corrected to the same ``global" power of \SI{22.5}{\watt}, obtaining the corrected signal $S_c$.  The applied correction is given by:
\begin{equation}
 S_c(f) = (S(f) - S_\text{BKG})\cdot C(f) + S_\text{BKG},
\end{equation}
where S$_\text{BKG}$ is the background level. The correction factors $C(f)$ are extracted by simulating the scaling ratio between the excitation probabilities at the effectively measured power and the global power for each frequency.

Because of the linear polarization of the field in the TLs, we can approximate our system to two levels and use the relevant optical Bloch equations \cite{2006_Haas} in the simulation. Single atom trajectories through both TLs reaching the Stop-MCP are simulated. While being in one of the TLs, the optical Bloch equations are numerically solved by an adaptive stepsize Runge-Kutta algorithm \cite{1992_RungeKutta}. To reproduce the field inside the TLs, for each TL a field map is generated in SIMION, following the procedure of Lundeen and Pipkin \cite{1986_Lundeen}. 
The numbers of excited and ground state atoms detected in the Stop-MCP are counted and from that the average excitation probability calculated. As input parameter to the simulation we assign to each atom an initial state with its specific resonance frequency, and the power for both TLs operating at random phases of the fields. The momentum and initial position of the specific particle from the LEM is simulated with the musrSIM package \cite{2012_musrSim} beforehand, from which the simulation randomly draws an atom. 

 The individual line-shape $P^{(i)}$, where $i$ stands for the assigned resonance frequency in \si{\mega\hertz}, is constructed by simulating the excitation probability for each initial state with large statistics over a frequency range between \SIrange{200}{2000}{\mega\hertz} in steps of \SI{1}{\mega\hertz}. Combining all relevant transitions, a global line-shape $P_n$ for a $n$ state is obtained:
\begin{align}
\label{eq:lineshape_equation}
    P_{n=2} &= 0.5\cdot P^{(1140)} + 0.25\cdot P^{(583)} + 0.25\cdot P^{(1326)}\\
    P_{n=3} &= 0.5\cdot P^{(339)} + 0.25\cdot P^{(174)} + 0.25\cdot P^{(394)},
\end{align}
where the constant factors are coming from spin statistics.

The fitting function is constructed with the simulated line-shapes:
\begin{equation}
\label{eq:fitting_equation}
    S_{c} = S_\text{BKG} + \sum_{n}\Big[B_{n}\cdot P_{n}\left(f - f_\text{offset}\right)\Big],
\end{equation}
where $B_n$ is scaling the line-shape $P_n$ for a specific $n$ state and $f_\text{offset}$ is introduced to allow for a global frequency offset compared to the theoretical resonance frequency used in the simulation.

\section{Results}
%
%
	\begin{figure}[t!]
    \includegraphics[width=1\columnwidth]{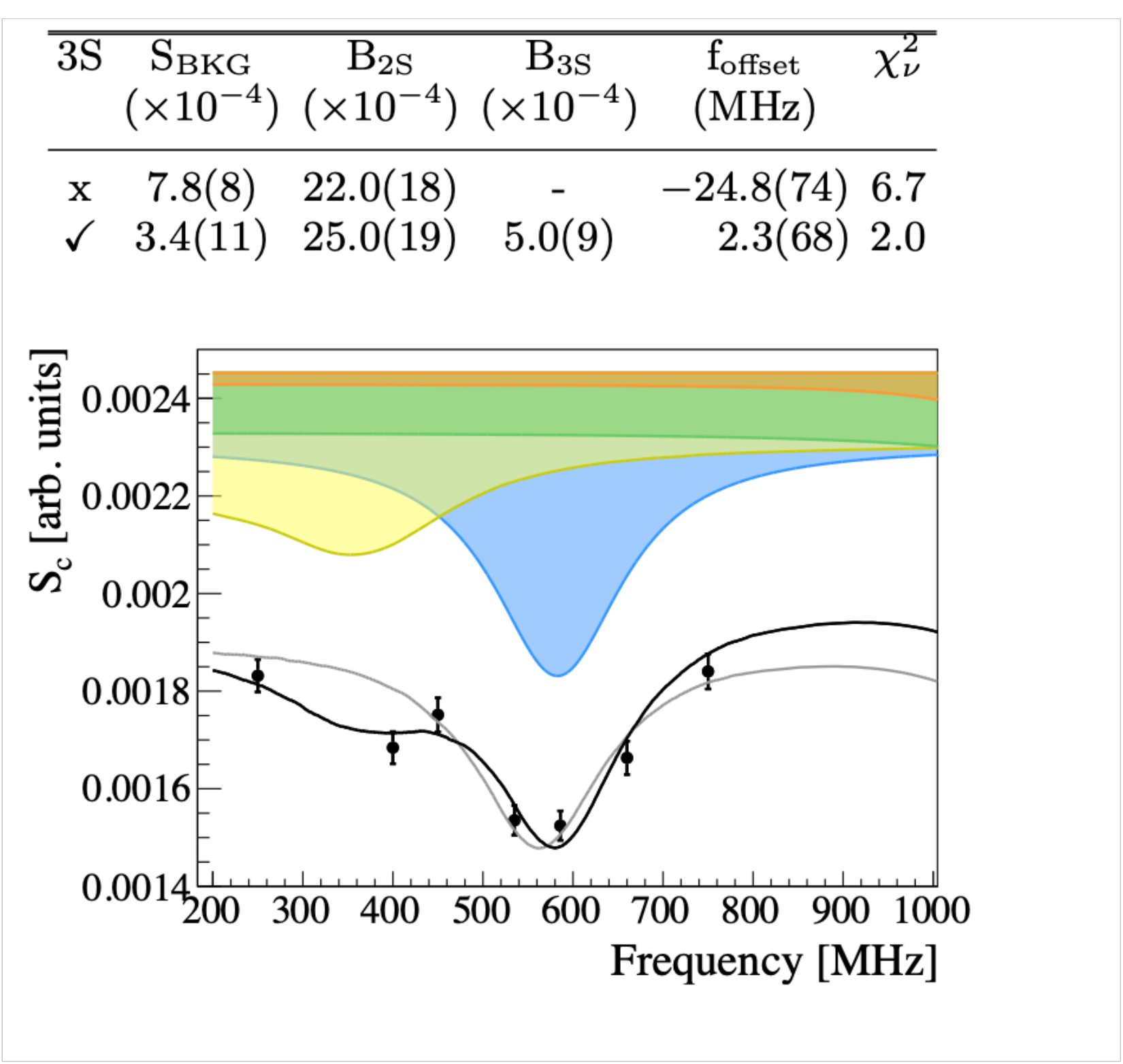}
    \caption{ \label{fig:3S}  Muonium scan at \SI{22.5}{\watt} in the range of \num{200} to \SI{800}{\mega\hertz}. The fitted black line is with, the gray line without the $3S$ contribution. The colored areas represent the underlying contributions from $2S-2P_{1/2}$ transitions, namely \SI{583}{\mega\hertz} (blue), \SI{1140}{\mega\hertz} (orange), \SI{1326}{\mega\hertz} (green), and the combined $3S-3P_{1/2}$ (yellow). The data point with TL OFF is not displayed in the figure, but is included in the fit; it lies at \num{20.4+-0.4 e-4}.}
 \end{figure}
 
The experimental data and the fits are shown in Fig.~\ref{fig:3S}. When the data is fit without any $3S$ contribution and hence $B_{3S}$ fixed to \num{0}, the reduced $\chi^2$ is \num{6.7} and one obtains a \SI{-24.8+-7.4}{\mega\hertz} offset compared to the theoretical value. By freeing the $3S$ population parameter, the fit improves to a reduced $\chi^2$ of \num{2.0}. The frequency offset is found to be \SI{2.3+-6.8}{\mega\hertz}. Both fits of the data are shown in Fig.~\ref{fig:3S}, where the gray line corresponds to the fit without and the black line with a $B_{3S}$ contribution. The colored line-shapes represent the underlying transitions with resonances at \SI{583}{\mega\hertz} (blue), \SI{1140}{\mega\hertz} (orange), \SI{1326}{\mega\hertz} (green) and a combined $3S-3P_{1/2}$ line-shape (yellow).

The main systematic uncertainties are similar to the ones we calculated in Ref.~\cite{2021_Ohayon} and total in \SI{0.19}{\mega\hertz}. The results are summarized in table \ref{tab:results}. The main difference is that the beam contamination in form of $3S$ states is taken into account in the fitting error already. Furthermore, due to their dependence on the resonance frequency, the systematic error stemming from the Doppler shift is approximately halved and the one coming from the uncertainty in the MW field intensity is doubled.

\begin{figure}[h!]
    \includegraphics[width=1\columnwidth,trim={90 5 30 30},clip]{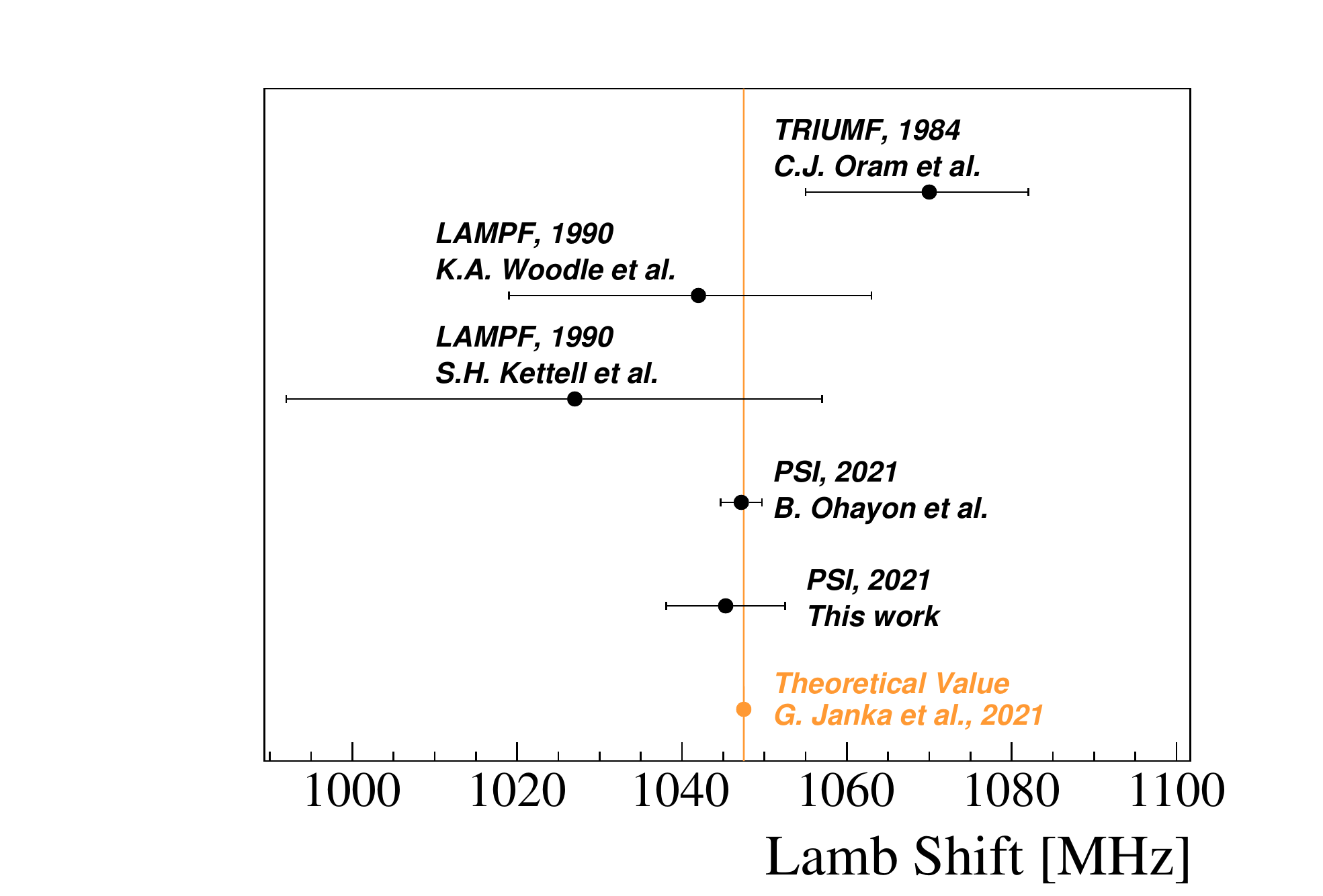}
    \caption{ \label{fig:LS_summary} Summary of all measurements of the $n=2$ Lamb shift in Muonium (black) \cite{1984_Oram, 1990_Woodle, 1990_Kettell, 2021_Ohayon} and the latest theoretical value (orange) \cite{2021_Janka}.}
        \end{figure}

From this measurement we extract the $2S_{1/2},F$=$0 \rightarrow 2P_{1/2},F$=$1$ resonance frequency to be \SI{580.6+-6.8}{\mega\hertz} and determine the M LS to be \SI{1045.5+-6.8}{\mega\hertz}. Our result agrees well with the theoretical value and is limited by the statistical uncertainty. A summary of all available measured values of the M LS is shown in Fig.~\ref{fig:LS_summary}. Using our previous results of the $2S_{1/2},F$=$1 \rightarrow 2P_{1/2},F$=$1$ resonance frequency \cite{2021_Ohayon}, we extract for the first time the $2S$ hyperfine splitting in Muonium to be \SI{559.6+-7.2}{\mega\hertz}. 

\begin{table}[h!]

    \begin{ruledtabular} 
\begin{tabular}{lcc}
                & Central Value & Uncertainty\\
    Fitting              &  $\hphantom{0}580.2\hphantom{00}$      &$\hphantom{<}6.8\hphantom{00}$  \\
    MW-Beam alignment    &        & $<0.16\hphantom{0}$ \\
    MW field intensity      &        & $<0.07\hphantom{0}$  \\
    M velocity distribution          &        & $<0.01\hphantom{0}$  \\
    AC Stark $2P_{3/2}$      &~ $+0.39$ & $<0.02\hphantom{0}$ \\ 
    \nth{2}-order Doppler  &~ $+0.03$ & $<0.01\hphantom{0}$ \\ 
    Earth's Field          &        & $<0.05\hphantom{0}$ \\ 
    Quantum Interference   &        & $<0.04\hphantom{0}$ \\ 
\noalign{\vspace{2pt}}
 \hline
\noalign{\vspace{2pt}}
    $2S,F$=$0$$-$$2P_{1/2},F$=$1$  & $\hphantom{0}580.6\hphantom{00}$  &$\hphantom{<}6.8\hphantom{00}$ \\ 
 \hline
 \noalign{\vspace{4pt}}

        Lamb Shift & $1045.5\hphantom{00} $ & $\hphantom{<}6.8\hphantom{00}$ \\        
        Theoretical value LS \cite{2021_Janka} &$1047.498$  &$\hphantom{<}0.001$ \\ 
\noalign{\vspace{2pt}}
 \hline
\noalign{\vspace{2pt}}

        2S HFS & $\hphantom{0}559.6\hphantom{00}$ &$\hphantom{<}7.2\hphantom{00}$  \\ 
        Theoretical value HFS (*) &$\hphantom{0}557.9\hphantom{00}$  &$<0.1\hphantom{00}$ \\ 
    \end{tabular}
    \end{ruledtabular}
    \begin{flushleft}* 
    The value was estimated by taking the most precise $1S$ HFS measurement in muonium \cite{1999_Liu} and dividing it by a factor of \num{2}$^3$. The uncertainty was estimated by calculating the QED $D_{21}$ difference \cite{2002_Karshenboim}. 
  \end{flushleft}
    \caption{\label{tab:results}Central values and uncertainty contributions in MHz.}
\end{table}

The detected $B_{3S}/B_{2S}$ ratio is \num{0.20+-0.04}. In the \SI{60}{\nano\second} from the foil to the entrance of the detection setup for an average M atom, \SI{32}{\percent} of the $3S$ states relax back to the ground state. Therefore, the effectively detected amount of $3S/2S$ created at the foil is \num{0.29+-0.07}. This agrees with the estimations done by C.~Fry \cite{1985_Fry} of \num{0.36} and is slightly lower than the estimate of \num{0.44+-0.04} from a combination of hydrogen population measurements \cite{1981_Gabrielse, 1973_Tielert}. 
An additional uncertainty in the estimated $3S/2S$ ratio could arise from the assumption that the detection efficiency of Ly$\alpha$ and Ly$\beta$ is similar. In fact, the efficiency depends on a number of factors, which make an accurate determination of its wavelength-dependency challenging (see Ref.~\cite{1987_siegmund,2005_Siegmund}).

\section{Conclusions}
To conclude, we measured for the first time the $2S_{1/2},F$=$0 \rightarrow 2P_{1/2},F$=$1$ transition in Muonium. This is the same hyperfine transition as the one used for the most precise determination of the hydrogen Lamb shift \cite{2019_Hessels}. Since this transition is even more isolated from the other hyperfine transitions in Muonium compared to hydrogen, this is the best candidate for future precision M LS measurements. By combining the data set with the scan of the \SI{1140}{\mega\hertz} resonance and leaving the $2S$ HFS splitting as a free parameter, we extract the $2S$ hyperfine splitting in Muonium also for the first time.
Furthermore, we detected M(3S) for the first time. This observation opens up the possibility for new microwave spectroscopy experiments with M such as the $n=3$ LS or the two-photon transition $3S-3D_{5/2}$. Both these transitions were measured in H to a high precision \cite{1980_Pipkin3S3D, 1972_Pipkin}.

In our measurement, the comparably large population of $3S$ states distorts the line-shape and introduces line-pulling, which might seem to defeat the purpose of choosing the isolated $n=2$ transition. However, the $n=3$ population can be electrically quenched with a weak electrical field, leaving a large fraction of the $n=2$ population unharmed as shown by measurements from Bezginov et al.~\cite{2020_Bezginov}. Extending the beamline to depopulate the $3S$ due to its lifetime is another option, but the strong beam straggling at the foil and resulting diffuse beam would reduce the $2S$ flux significantly. This problem could be overcome by exchanging the carbon foil with only a few layers of graphene (around \SI{1}{\nano\meter} thickness) \cite{2014_Allegrini} or moving to a gaseous target. Additionally, with a new apparatus such as muCool to compress and cool the beam, or the HiMB project to increase the $\mu^+$ flux at PSI, the measurements would not be statistically limited anymore. This would allow to probe QED effects such as the Barker-Glover and the nucleus self-energy, which are not accessible in hydrogen yet \cite{2021_Janka}. 

\section*{Acknowledgements} 
This work is based on experiments performed at the Swiss Muon Source S$\mu$S, Paul Scherrer Institute, Villigen, Switzerland.
This work is supported by the ERC consolidator grant 818053-Mu-MASS and the Swiss National Science Foundation under the grant 197346.
BO acknowledges support from the European Union’s Horizon 2020 research and innovation program under the Marie Skłodowska-Curie grant agreement No.~101019414, as well as ETH Zurich through a Career Seed Grant SEED-09 20-1.

\bibliographystyle{apsrev4-1}
\bibliography{sample.bib}

\end{document}